\documentclass[12pt]{iopart}

\usepackage{graphicx,bbold} 

\def\({\left(}
\def\){\right)}
\def\dA#1#2{\delta A^{(#1)}_{#2}}
\def\dB#1#2{\delta B^{(#1)}_{#2}}

\def\av#1{\left\langle #1\right\rangle}
\def\ad{a^\dagger}
\def\d#1{#1^\dagger}

\def\eq#1{Eq.~(\ref{eq:#1})}
\def\Eq#1{Equation~(\ref{eq:#1})}

\def\fig#1{Fig.~\ref{fig:#1}}
\def\Fig#1{Figure~\ref{fig:#1}}

\def\twocol#1#2{\left(\begin{array}{c} #1 \\ #2 \end{array}\right)}
\newcommand\fourcol[4]{\left(\begin{array}{c} #1 \\ #2\\ #3 \\ #4\end{array}\right)}
\def\twomat#1#2#3#4{\left(\begin{array}{ccc}
#1 & #2 \\ 
#3 & #4 \end{array}\right)} 

\begin{document}

\title[Nonlocal restoration of two-mode squeezing]{Nonlocal restoration of two-mode squeezing in the presence of strong optical loss}
\author{Russell Bloomer}
\author{Matthew Pysher}
\author{Olivier Pfister}
\ead{opfister@virginia.edu}
\address{Department of Physics, University of Virginia, Charlottesville, Virginia 22903, USA}


\date{\today}

\begin{abstract}
We present the experimental realization of a theoretical effect discovered by Olivares and Paris \cite{Olivares2009}, in which a pair of entangled optical beams undergoing independent losses can see nonlocal correlations restored  by the use of a nonlocal resource correlating the losses. Twin optical beams created in an entangled Einstein-Podolsky-Rosen (EPR) state by an optical parametric oscillator above threshold were subjected to 50\% loss  from  beamsplitters in their  paths. The resulting severe degradation of the signature quantum correlations observed between the two beams was then suppressed when another, independent EPR state impinged upon the other input ports of the beamsplitters, effectively entangling the losses inflicted to the initial EPR state. The additional EPR beam pair was classically coherent with the primary one but had no quantum correlations with it. This result may find applications as  a ``quantum tap'' for entanglement.
\end{abstract}

\maketitle

\section{Introduction}
\begin{figure}[hbt]
\centerline{
\includegraphics[width=\columnwidth]{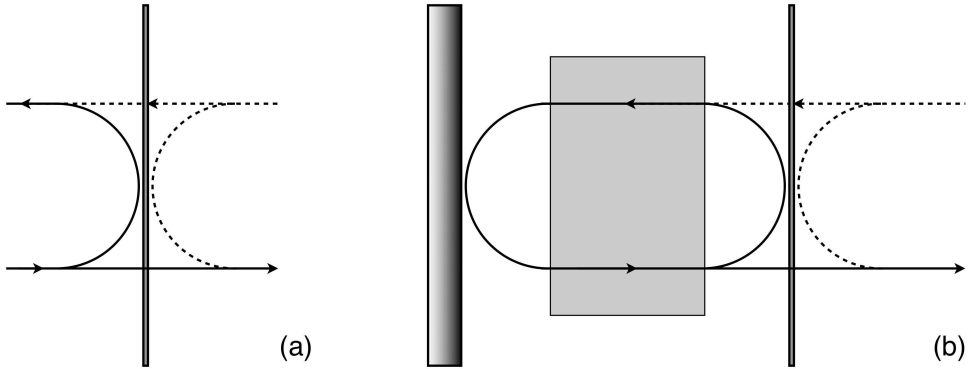}
}
\caption{(a) Decohering interaction studied here. A quantum signal impinges on to the lower left port of the beam splitter and we measure the transmitted (or reflected) beam. Both outputs are ``contaminated'' by vacuum fluctuations (upper right port) in a unitary, reversible manner. (b) Decoherence model with a cavity, the resonant modes of which are populated by an optical amplifier and by the vacuum modes entering through the output mirror on the right. Vacuum fluctuations therefore ``contaminate'' the quantum evolution of the light emitter round trip after round trip, in a way that becomes irreversible and necessitates tracing over the reservoir.}
\label{fig:decoh}
\end{figure}

Quantum decoherence is a major impediment to experimental advances in quantum optics and quantum information. It arises from the unitary coupling of the physical system of interest to an infinite reservoir, upon which an average must be taken in order to yield tractable predictions. This statistical average yields irreversible, measurement-like modifications of the quantum state that are incompatible with desired unitary, coherent quantum evolution. Here, we present experimental results on decoherence control. In order to best present our results, which pertain to strong loss rather than decoherence, it is useful to give a few basic reminders about the latter. A concrete and general example of a reservoir for boson fields is the infinite reservoir of vacuum modes which are coupled to the system by the Hamiltonian \cite{Louisell1973}
\begin{equation}
H_{c} = i\hbar \sum_j \,\kappa_j\,\left(a^\dagger v_j- a\,v^\dagger_j\right),
\label{eq:hc}
\end{equation}
where $a$ is the annihilation operator of a ``signal'' mode of the quantized electromagnetic field, $v_j$ are the annihilation operators of the vacuum modes (we take the reservoir to be a large but finite cavity of corresponding resonant frequencies $\omega_j$), and $\kappa_j$ is the coupling rate, chosen real, between the signal and vacuum mode $j$. This Hamiltonian describes all one-quantum energy exchanges (e.g.\ beamsplitter for a field, dipole electric coupling for an atom) between the signal and the reservoir, a situation depicted in \fig{decoh}(a). 
The quantum statistics arising from the interaction of signal $a$ with a \emph{single} vacuum mode $v$ at the same frequency (and polarization) were first derived by Caves in the context of quantum interferometry \cite{Caves1980}. The beam splitter yields the following unitary transformation
\begin{equation}
\twocol{a(t)}{ v(t)} = \twomat{\cos \kappa t}{-\sin \kappa t}{\sin \kappa t}{\cos \kappa t}\twocol{a(0)}{ v(0)}.
\label{eq:bounce2}
\end{equation}
which admixes the signal with vacuum fluctuations identical to that of a coherent state. In our experimental work, we'll maximize this effect by choosing $\kappa$ such that $\cos\kappa t=1/\sqrt 2$. This admixture is reversible since it results from unitary evolution. The simplest way to see that is to consider the SU(2) invariant, in this case, the total photon number:
\begin{equation}
a(t)^\dag a(t)+ v(t)^\dag v(t)=a(0)^\dag a(0)+v(0)^\dag v(0).
\label{eq:cons}
\end{equation}
\Eq{cons} clearly suggests that taking the sum of photodetected signals after a beam splitter will  restore the photon number statistics of input signal $a(0)$, with the vacuum input playing no role in this case. This is a well known quantum optical  detection technique.

Now, consider one signal mode and $N$ vacuum modes. The Heisenberg equation for interaction (\ref{eq:hc}) becomes
\begin{equation}
\frac d{dt}\fourcol{a}{ v_1}\vdots{ v_N} = 
\left(\begin{array}{cccc} 
0 & \kappa_1 & \ldots & \kappa_N \\
-\kappa_1 & 0 & \ldots & 0\\
\vdots &  \vdots & \ddots & \vdots\\
-\kappa_N & 0 & \ldots & 0
\end{array}\right)
\fourcol{a}{ v_1}\vdots{v_N},
\end{equation}
which is straightforward to solve: out of the $N$+1 initial modes, only two aren't constants of motion and undergo temporal evolution. These modes are $a$ and the ``supermode'' 
\begin{equation}
\mathcal V=\sum_j \frac{\kappa_j}g\,v_j,
\label{eq:supermode}
\end{equation}
where $g=(\sum_j\kappa_j^2)^{1/2}$. The solutions are
\begin{equation}
\twocol{a(t)}{\mathcal V(t)} = \twomat{\cos gt}{-\sin gt}{\sin gt}{\cos gt}\twocol{a(0)}{\mathcal V(0)}.
\label{eq:bounce}
\end{equation}
One should note the difference between \eq{bounce} and \eq{bounce2}: for uniform coupling strengths $\kappa_j=\kappa$ across the reservoir, $g=\kappa\sqrt N$, i.e.\ $g$ grows with the reservoir's bandwidth. The characteristic time $t_c$\,$\sim$\,$g^{-1}$ that leads to significant admixing of $a$ with vacuum supermode $\mathcal V$ therefore becomes extremely small when the size of the reservoir increases. This evolution is still unitary, and therefore still reversible, even though it now involves a broadband mode distribution.

In systems such as a laser, the light is emitted inside a cavity and \eq{bounce} describes but a single instance of the coupling of the laser field out of the laser cavity, see \fig{decoh}(a). Because the cavity storage time is longer than $t_c$, many more such instances occur on average (\fig{decoh}(b)). The  standard model of decoherence is then the well-known Markovian treatment, carried out in a coarse-grained Heisenberg-Langevin description to yield the fluctuation-dissipation theorem \cite{Louisell1973}
\begin{equation}
\frac{d a}{d t} = -\left(\frac\Gamma2+i\Delta\right)  a + F, 
\label{dec}
\end{equation}
where $\Gamma$ is the decay rate, $\Delta$ a frequency shift (of which the Lamb shift is an example for atoms \cite{Sargent1974}), and $F$ the quantum Langevin fluctuation made up by the vacuum modes. 

Here, we focus on the experimental situation of \fig{decoh}(a) with a 50\% reflecting beam splitter. Only the transmitted beam will be measured, i.e.\ there will be a 50\% loss at all times, the reflected beam being considered as irretrievably lost. However, we will claim the option of accessing both input ports, which still corresponds to realistic and feasible, if not universal, experimental situations.

\subsection{Model system}
\begin{figure}[hbt]
\centerline{
\includegraphics[width=\columnwidth]{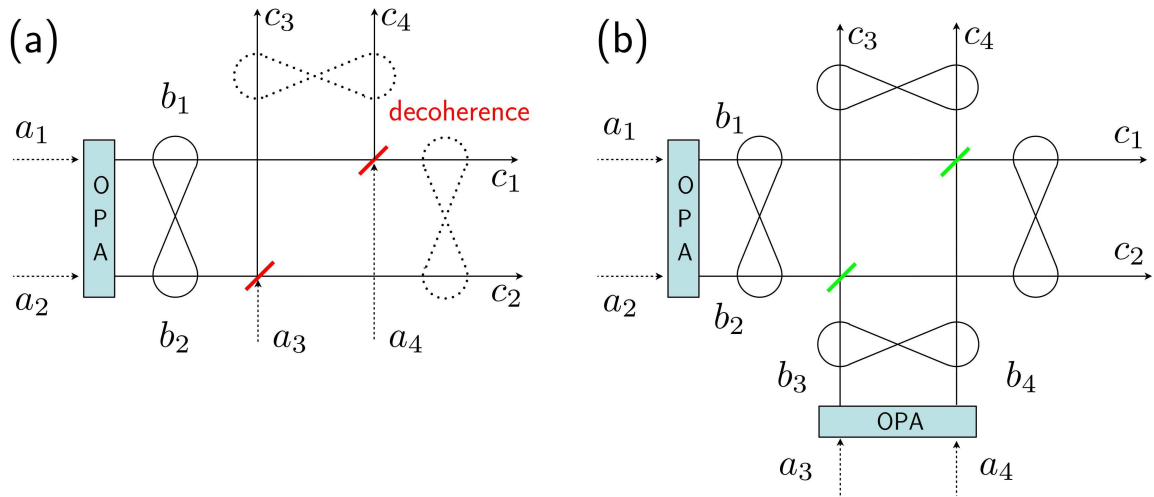}
}
\caption{Model for loss decoherence of an EPR state. (a), loss decoherence created by two 50\% beam splitters. (b), Olivares and Paris' proposal for decoherence protection of an EPR state \cite{Olivares2009}. All input modes $a_{1-4}$ are in a vacuum state, and the two OPAs are independent.}
\label{fig:OP}
\end{figure}

In the theoretical proposal by Olivares and Paris \cite{Olivares2009}, a nondegenerate optical parametric oscillator (OPO) below threshold generates an optical Einstein-Podolsky-Rosen (EPR) entangled state \cite{Einstein1935}, a pair of ``twin'' beams emitted  by spontaneous parametric downconversion and filtered by the doubly resonant OPO cavity \cite{Reid1989,Ou1992}. The twin EPR beams  are amplitude-correlated and phase-anticorrelated, i.e.\ exhibit quantum noise reduction (squeezing) of the amplitude difference and phase sum, and are  subjected to decoherence by traversing 50\%-reflectivity beamsplitters, \fig{OP}(a). 
The independent fluctuations of the vacuum modes $a_{3,4}$ that enter the other input ports of the beamsplitters are responsible for a drastic reduction of the initial EPR quantum correlations. Olivares and Paris showed that this decohering effect could be suppressed, not by preventing the beamsplitter interactions from occurring at all, but by replacing the vacuum incident on the beamsplitters with another, independent EPR state, thereby entangling the reservoirs at each distinct beamsplitter. In this sense, the reservoir fields in the unitary short-term couplings of  \eq{bounce}---one for each beamsplitter---become correlated so as to preserve the quantum correlations between each signal. In addition, this proposal offers a ``quantum tap'' for entanglement, as a direct generalization of that previously realized for single-mode squeezed states \cite{Bruckmeier1997}. 

As mentioned above, we do not address decoherence as a coarse-grained process but instead consider the shorter time scale of a single, unitary beamsplitter bounce. However, there is no fundamental limitation to extending this procedure to the coarse-grained time scale---injecting squeezing directly into the OPOs through their output mirrors---since we are dealing with the exact same Hamiltonian (\ref{eq:hc}). This interesting project is out of the scope of this paper \footnote{It was predicted that the Schawlow-Townes emission linewidth of a laser can be reduced by a factor of 2 if an infinitely squeezed vacuum field is injected into the cavity \protect\cite{Gea-Banacloche1987}.}. 

Besides Ref.~\cite{Olivares2009}, this work also relates directly to theoretical results on quantum channel capacity \cite{Cerf2005, Filip2005}, of which the work described in the following is, to our knowledge, the first experimental realization. Note that other proposals were made for quantum engineering decoherence \cite{Poyatos1996,Diehl2008}. Also, experimental generation of entanglement from decoherence was recently achieved \cite{Kraute2010}.

\subsection{Experimental setup}

\begin{figure}[hbt]
\centerline{\includegraphics[width=\columnwidth]{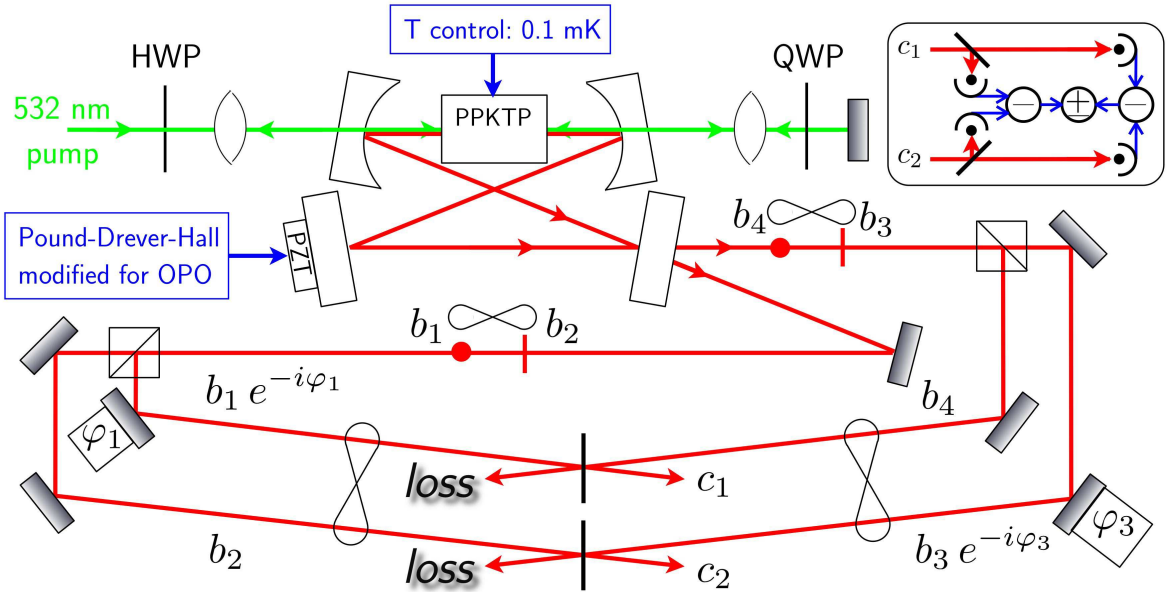}} 
\caption{Sketch of the experimental setup. A stabilized ring OPO above threshold downconverted individual photons from a continuous-wave pump field at a wavelength of 532 nm into crosspolarized photons pairs, yielding EPR beams of 1-10 mW continuous-wave power at wavelengths around 1064 nm. The EPR modes were separated by polarizing beamsplitters and phase shifted by piezotransducer-actuated mirrors. The initial pump beam's polarization was controlled by a half-wave plate (HWP) in order to set the power of EPR fields $b_{3,4}$. The counterpropagating pump beam was mode-matched to the initial one by a cat's eye, and a doubly passed quarter-wave plate (QWP) controlled its polarization in order to set the power of EPR fields $b_{1,2}$. Inset top right, detailed detection network. The ``+'' output gives the quantum signal, the final ``-'' output gives the shot noise.
}
\label{fig:setup}
\end{figure}

In our experiment, the entanglement was generated by an OPO above threshold \cite{Reid1988,Villar2005,Su2006,Jing2006,Keller2008}. The quantum correlations are of the same nature as those below the threshold, but the predominant emission mechanism is stimulated instead of spontaneous. Our experimental implementation is depicted in Fig. \ref{fig:setup}. The OPO  was constituted of a ring resonator containing a nonlinear periodically poled $\rm KTiOPO_4$ crystal, which was pumped by 
two counterpropagating pump beams of frequency $\omega_p$, emitted at 532 nm by a stable continuous-wave frequency-doubled Nd:YAG laser (Innolight Diabolo). The OPO thereby emitted two counterpropagating independent pairs of continuous-wave, orthogonally polarized EPR fields ($b_1$,$b_2$) and ($b_3$,$b_4$). The OPO cavity length was stabilized by a variant of the Pound-Drever-Hall technique \cite{Drever1983,Feng2004} and each EPR pair thus had well-defined, if arbitrary, individual beam frequencies ($\omega_H$,$\omega_V$) such that $\omega_H$+$\omega_V$=$\omega_p$. Because the counterpropagating pump beams were carefully mode-matched, they excited the same OPO mode pair and $\omega_V$ was identical for $b_1$ and $b_4$ in each EPR pair (same with $\omega_H$ for $b_2$ and $b_3$). This property was crucial to the success of the experiment since the interfering fields at a beamsplitter must be indistinguishable for the appropriate multimode quantum evolution \cite{Olivares2009} to take place, as epitomized in the Hong-Ou-Mandel effect \cite{Hong1987, Feng2004}. 
After separation and phase shifting, the frequency-degenerate EPR beams from each pair were overlapped at the ``loss'' beamsplitters. Intensity measurements were then performed by high quantum efficiency photodiodes (JDSU ETX500T) on only a {single} output of each beamsplitter, the other output being treated as an irretrievable loss channel (Fig. \ref{fig:setup}). Note that these measurements only allowed us to monitor the intensity-difference (i.e.\ amplitude-difference in the linear approximation) squeezing. Although true EPR entanglement necessitates phase-sum squeezing as well, we'll see in the next section that our  intensity-difference measurements results do prove unequivocally the existence of EPR correlations at the input, even without any phase-sum measurement.

\section{Theory}

The complete theoretical analysis can be found for pure EPR states in \cite{Olivares2009}. Here, we give a linearized theory in the Heisenberg picture, in order to describe our particular (and different) experimental situation of an OPO above threshold.

\subsection{Quantum fluctuations of the output intensity difference}
The outputs from one side of the beamsplitters are
\begin{eqnarray}
c_1 &=& \frac1{\sqrt2}\(b_1\,e^{i\varphi_1}+b_4\,e^{i\varphi_4}\),  \\
c_2 &=& \frac1{\sqrt2}\(b_2\,e^{i\varphi_2}+b_3\,e^{i\varphi_3}\).
\end{eqnarray}
As mentioned above, the output fields on the other side of the beamsplitters are ``lost'' and will never be used in the theory or the experiment.
The linearized output field operators, \fig{setup}, are ($\beta_1=\beta_2$, $\beta_3=\beta_4$)
\begin{eqnarray}
c_1 &=& \gamma_1+\delta c_1\ \equiv\ 
\frac1{\sqrt2}\(\beta_1\,e^{i\varphi_1} +\beta_3\,e^{i\varphi_4}\) +\frac1{\sqrt2}\(\delta b_1\,e^{i\varphi_1}+\delta b_4\,e^{i\varphi_4}\),\\
c_2 &=& \gamma_2+\delta c_2\ \equiv\ 
\frac1{\sqrt2}\(\beta_1\,e^{i\varphi_2} +\beta_3\,e^{i\varphi_3}\) +\frac1{\sqrt2}\(\delta b_2\,e^{i\varphi_2}+\delta b_3\,e^{i\varphi_3}\).
\end{eqnarray}
Note that the classical interference fringes are given by
\begin{eqnarray}
n_1 &=& |\gamma_1|^2 = \frac{\beta_1^2+\beta_3^2}2 +\beta_1\beta_3\cos\varphi_{14}, \label{eq:class1}\\
n_2 &=& |\gamma_2|^2 = \frac{\beta_1^2+\beta_3^2}2 +\beta_1\beta_3\cos\varphi_{23}.
\label{eq:class2}
\end{eqnarray}
Moving on to the linearized intensity fluctuations, we recall that fields $c_{1,2}$ are both the sums of two interfering fields so we must keep track of individual phases. We use the generalized quadrature fluctuations defined by
\begin{equation}
\dA{j}\theta = \frac1{\sqrt2}\(\delta a_j \,e^{-i\theta}+\delta \ad_j\, e^{i\theta}\),
\end{equation}
where the values $\varphi$=0,$\pi/2$ denote the amplitude and phase quadratures, respectively. We also use
\begin{eqnarray}
N&=&\d cc=(\gamma^*+\delta\d c)(\gamma+\delta c)=|\gamma|^2+\(\gamma^*\,\delta c+\gamma\,\delta\d c\) + \delta\d c\delta c,\\
\delta N  &\simeq& \gamma^*\,\delta c+\gamma\,\delta\d c,
\end{eqnarray}
so that 
\begin{eqnarray}
\delta(N_{1} - N_2)&=& \frac{\beta_1}{\sqrt2} \(\dB10-\dB20-\dB3{\varphi_{23}}+\dB4{\varphi_{14}}\)\nonumber\\
&&+\frac{\beta_3}{\sqrt2}\(\dB1{-\varphi_{14}}-\dB2{-\varphi_{23}}-\dB30+\dB40\),
 \label{eq:dN}
\end{eqnarray}
where $\varphi_{ij}=\varphi_i-\varphi_j$. From this, we unsurprisingly see that only the relative phases at the beam splitters matter. These phases were easily controlled experimentally by tuning $\varphi_1$ and $\varphi_3$ using piezo-actuated mirrors.

\subsection{Squeezed signals}

We first consider the case of \fig{OP}(a), where $b_{3,4}$ are only vacuum fields. In this case, the variance $V(N_{1} - N_2) = \av{[\delta(N_{1} - N_2)]^2}$ will contain constant noise terms $\Delta B^{(3,4)}_\varphi=\av{[\dB{3,4}\varphi]^2}=1$ which  reduce the observed squeezing from mode pair (1,2).

Let's now examine the conditions for the output intensity difference to be maximally squeezed. The first straightforward case is to assume 2 pairs of EPR beams, i.e.
\begin{eqnarray}
\dB10\pm\dB20 &=& \(\dA10\pm\dA20\)\,e^{\pm r}, \\
\dB1{\frac\pi2}\pm\dB2{\frac\pi2} &=& \(\dA1{\frac\pi2}\pm\dA2{\frac\pi2}\)\,e^{\mp r}, \\
\dB30\pm\dB40 &=& \(\dA30\pm\dA40\)\,e^{\pm s}, \\
\dB3{\frac\pi2}\pm\dB4{\frac\pi2} &=& \(\dA3{\frac\pi2}\pm\dA4{\frac\pi2}\)\,e^{\mp s}. 
\end{eqnarray}
From this we see that \eq{dN} will give squeezing for ($\varphi_{14}$,$\varphi_{23}$) = (0,0), ($\pi$,$\pi$), and ($\pm\pi/2$,$\mp\pi/2$). Conversely, the worst possible cases are an admixture of squeezing and antisqueezing for (0,$\pi$), ($\pi$,0), and ($\pm\pi/2$,$\pm\pi/2$). 

One might also want to ask what would happen should only single-mode amplitude squeezing be present. In that case, it is easy to see that the prediction for ($\varphi_{14}$,$\varphi_{23}$) = (0,0) or ($\pi$,$\pi$) is indistinguishable from the EPR case. However, such is not the case for ($\varphi_{14}$,$\varphi_{23}$) =  ($\pm\pi/2$,$\mp\pi/2$), since it is impossible to simultaneously squeeze the amplitude and phase quadratures. For the latter phase value set, \emph{only genuine EPR correlations can give a squeezed signal on the output intensity difference}. This therefore amounts to an indirect measurement of the EPR paradox at the beamsplitters' input.

\subsection{Generalized quadrature squeezing}

The aforementioned particular EPR cases are not the only ones possible. If we express the photon number fluctuations in terms of the input vacuum modes, we obtain, from the Bogoliubov equations
\begin{eqnarray}
\dB{1,2}\varphi &=& \cosh r\, \dA{1,2}\varphi +\sinh r\, \dA{2,1}{-\varphi}, \\
\dB{3,4}\varphi &=& \cosh s\, \dA{3,4}\varphi +\sinh s\, \dA{4,3}{-\varphi}, 
\end{eqnarray}
where $r$ and $s$ are the respective squeezing parameters of each EPR pair. It ensues that
\begin{eqnarray}
\delta(N_{1} - N_2)
&=&\quad \frac{\beta_1}{\sqrt2}\left[\(\delta A^{(1)}_{0}-\delta A^{(2)}_{0}\)\,e^{-r}
 - \delta A^{(3)}_{\varphi_{23}} \cosh s +\delta A^{(3)}_{-\varphi_{14}}\sinh s \right.\nonumber\\
 &&\quad\qquad\qquad\qquad\qquad\qquad\left. + \delta A^{(4)}_{\varphi_{14}}\cosh s - \delta A^{(4)}_{-\varphi_{23}}\sinh s\right]
 \nonumber\\ 
 && +\frac{\beta_3}{\sqrt2}\left[
\ \delta A^{(1)}_{-\varphi_{14}}\cosh r - \delta A^{(1)}_{\varphi_{23}}\sinh r \right.\nonumber\\
&&\left.\quad\qquad-\delta A^{(2)}_{-\varphi_{23}}\cosh r +\delta A^{(2)}_{\varphi_{14}}\sinh r
-\(\delta A^{(3)}_{0}-\delta A^{(4)}_{0}\)\,e^{-s} \right], \nonumber\\
\end{eqnarray}
from which it is clear that the general condition for number-difference squeezing at the measured outputs of the beam splitters is simply
\begin{equation}
\varphi_{14}=-\varphi_{23} 
= \varphi,
\end{equation}
(to $2\pi$ left). Indeed, in this case,
\begin{eqnarray}
\delta(N_{1} - N_2)
&=&\quad \frac{\beta_1}{\sqrt2}\left[\(\delta A^{(1)}_{0}-\delta A^{(2)}_{0}\)\,e^{-r}
 - \(\delta A^{(3)}_{-\varphi_{}}-\delta A^{(4)}_{\varphi_{}}\) e^{-s}\right]
 \nonumber\\ 
 && +\frac{\beta_3}{\sqrt2}\left[
 \(\delta A^{(1)}_{-\varphi_{}}-\delta A^{(2)}_{\varphi_{}}\) e^{ -r}
-\(\delta A^{(3)}_{0}-\delta A^{(4)}_{0}\)\,e^{-s} \right]. 
\label{eq:final}
\end{eqnarray}
In other words, even though the original EPR correlations appear explicitly for $\varphi=0,\pi$ (amplitude-difference) and $\varphi=\pm\pi/2$ (phase-sum), it is remarkable that $\varphi$ is in no way restricted to these particular values! \footnote{A similar effect was very recently reported for the interference of two single squeezed modes to generate an EPR state, by Leuchs et al. \cite{Leuchs2009}.} 

\subsection{Variance of the photon-number difference}
We can now derive the actual quantum noise
\begin{eqnarray}
V(N_1-N_2) 
&=& \ \ \frac{\beta_1^2}2\left[e^{-2r}+\cosh 2s - \sinh 2s\,\cos\(\varphi_{14}+\varphi_{23}\) \right] \nonumber\\
&&+\frac{\beta_3^2}2\left[\cosh 2r - \sinh 2r\,\cos\(\varphi_{14}+\varphi_{23}\)+e^{-2s} \right] \nonumber\\ 
&&  +\frac{\beta_1\beta_3}2\(e^{-2r}+e^{-2s}\)\(\cos\varphi_{14}+\cos\varphi_{23}\)  
\end{eqnarray}
and proceed to apply this expression to predicting our experimental results. The shot noise level is obtained for $r=s=0$, in which case
\begin{eqnarray}
V_\mathit{SN}
&=& {\beta_1^2}+{\beta_3^2}+{\beta_1\beta_3}\(\cos\varphi_{14}+\cos\varphi_{23}\).  
\end{eqnarray}
Unsurprisingly, the shot noise follows the classical interference fringes of \eq{class1} and \eq{class2}. The situation that interests us is $\varphi_{14}=-\varphi_{23}= \varphi$, in which case
\begin{eqnarray}
V(N_1-N_2) 
&=& \frac {e^{-2r}+e^{-2s}}2\({\beta_1^2}+{\beta_3^2}+2{\beta_1\beta_3}\cos\varphi\), \\
\label{eq:sqphi}
V_\mathit{SN}
&=& {\beta_1^2}+{\beta_3^2}+2{\beta_1\beta_3}\cos\varphi. 
\label{eq:snphi}
\end{eqnarray}
This shows that the signal is always squeezed by the compound factor $(e^{-2r}+e^{-2s})/2$ for all values of $\varphi$! We will later present the experimental test of these predictions.

\section{Experimental results}

\subsection{Vacuum inputs}

\begin{figure}[hbt]
\begin{center}
\parbox{0.5\columnwidth}{
\includegraphics[width=0.5\columnwidth]{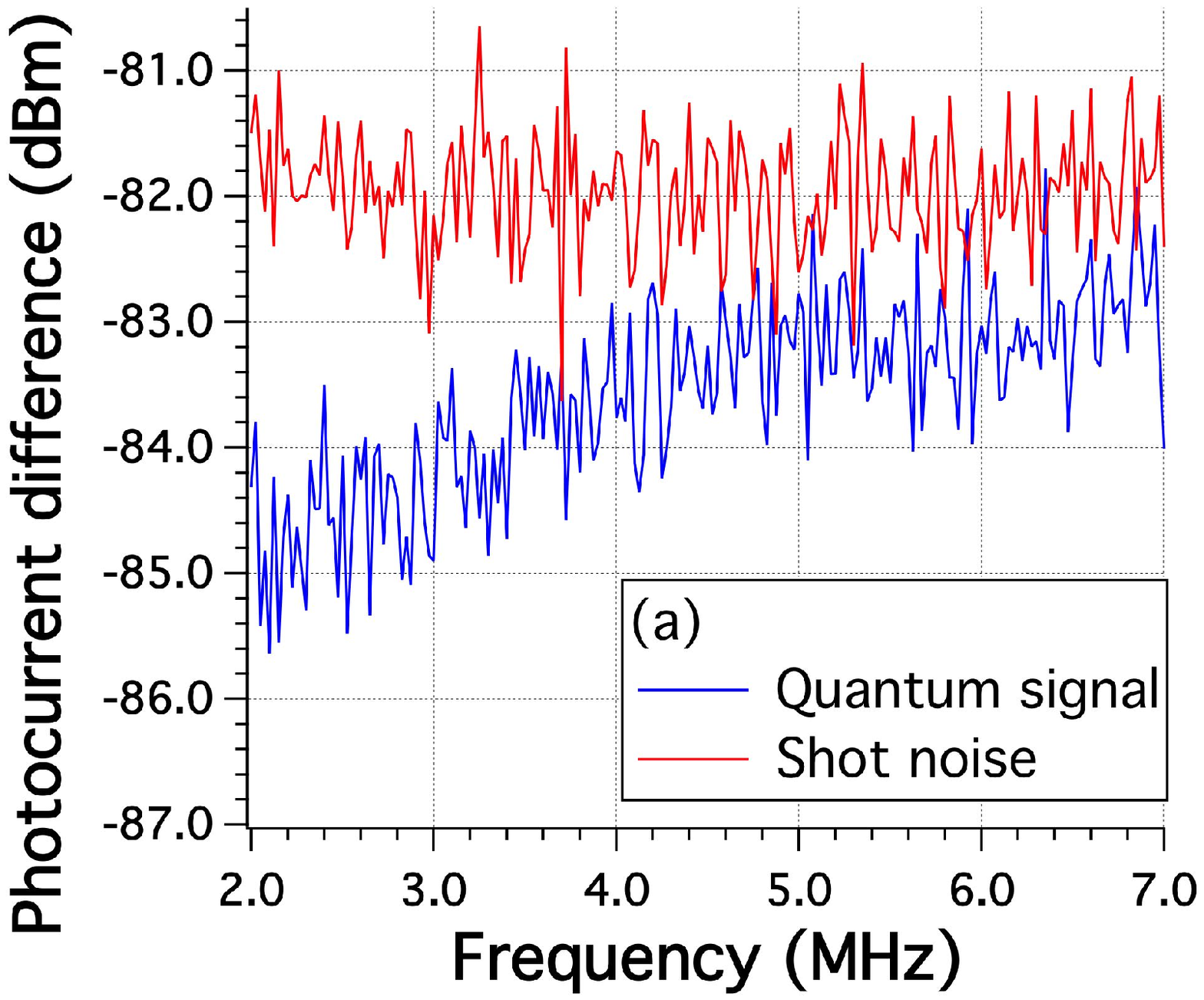}
}\parbox{0.5\columnwidth}{
\includegraphics[width=0.5\columnwidth]{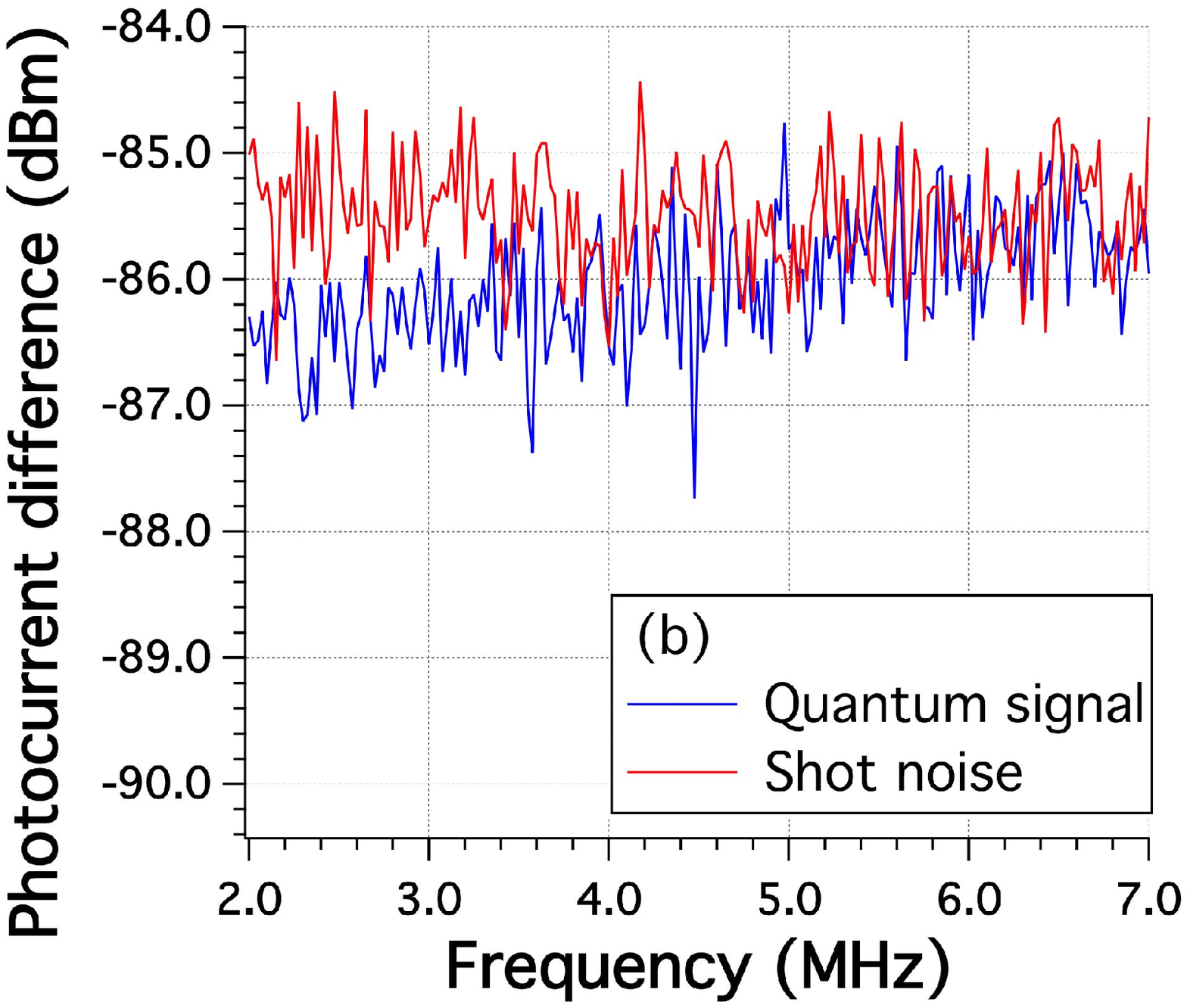}
}
\caption{Effect of the decoherence induced by a 50\% loss rate from two beamsplitters on an EPR state. All traces are power spectra of photocurrent differences of the two EPR beams. Their low-pass aspect solely reflects the OPO cavity bandwidth and is not an effect of experimental imperfections. (a) power spectrum of the photocurrent difference before the beamsplitters. (b) same, after the beamsplitters. (Resolution bandwidth: 30 kHz. Video bandwidth: 30 kHz. 100 averages.)}
\label{fig:sq}
\end{center}
\end{figure}

As detailed in the theory, we restricted ourselves to intensity-difference measurements of output modes $c_{1,2}$. When only OPO beams (1,2) were used, with modes (3,4) being vacuum, a severe degradation of the the initial intensity-difference squeezing was observed, as seen in \fig{sq}. 

\subsection{OPO inputs}

\begin{figure}[hbt]
\begin{center}
\parbox{0.5\columnwidth}{
\includegraphics[width=0.5\columnwidth]{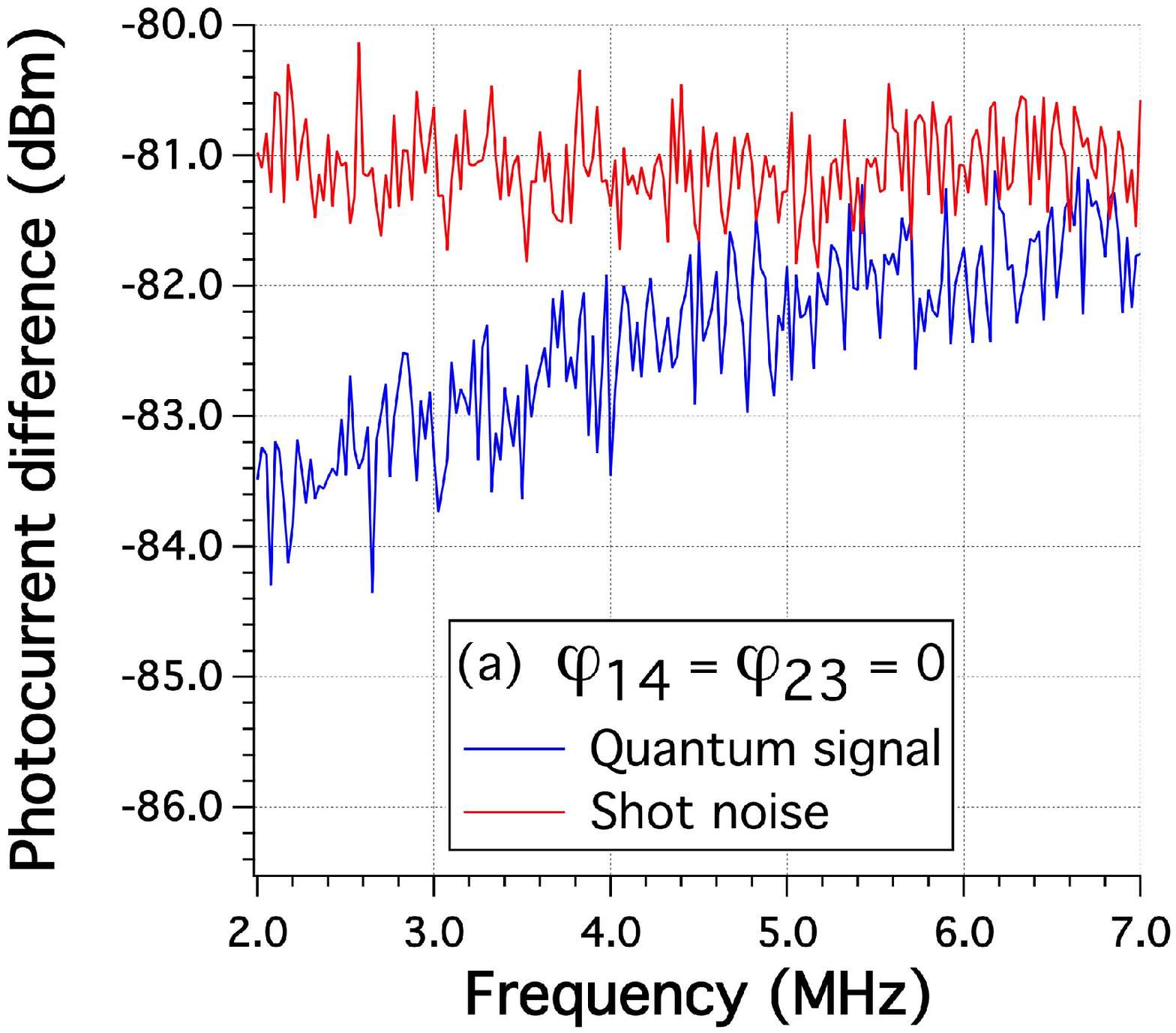}
}\parbox{0.5\columnwidth}{
\includegraphics[width=0.5\columnwidth]{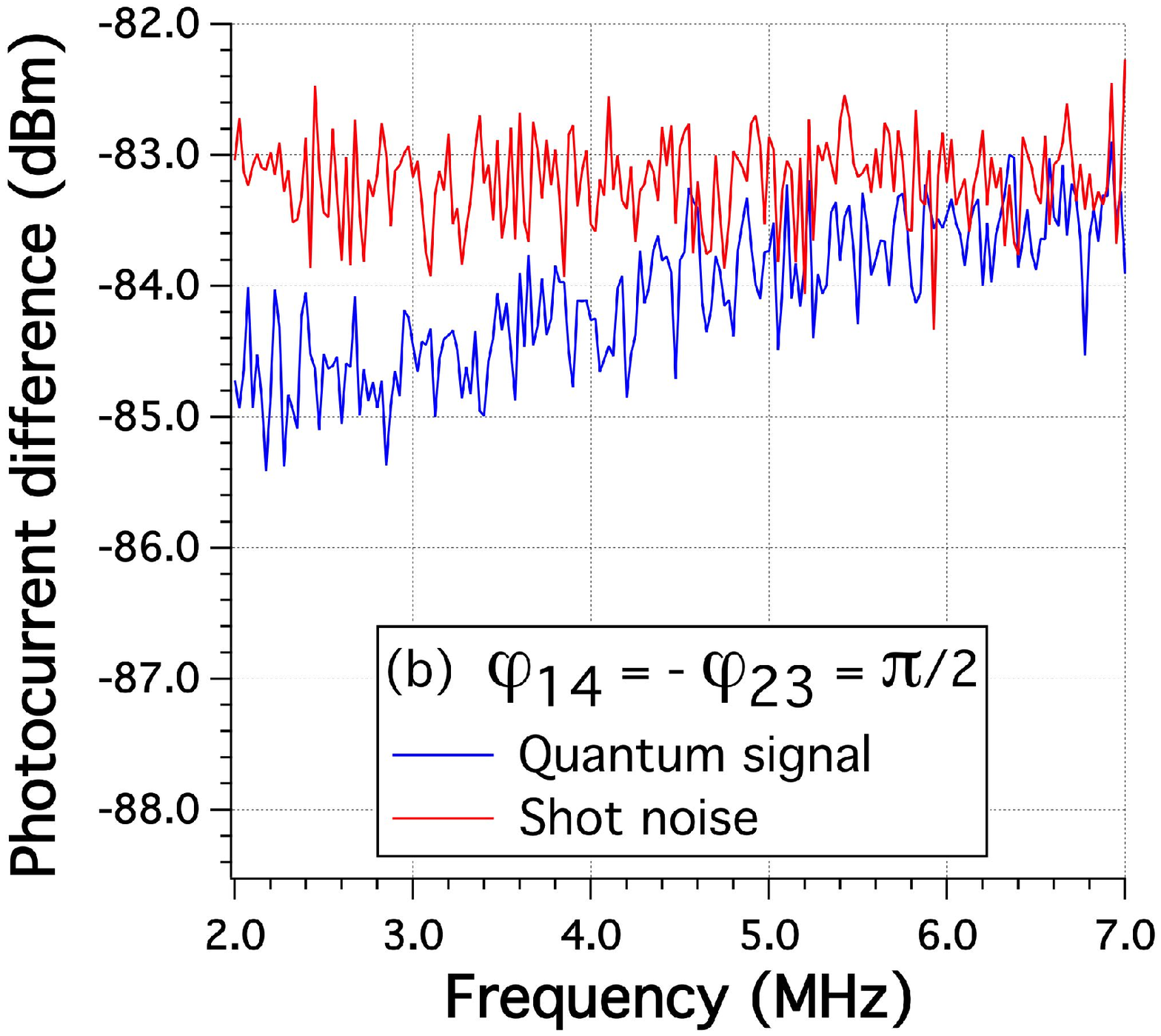}
}
\parbox{0.5\columnwidth}{
\includegraphics[width=0.5\columnwidth]{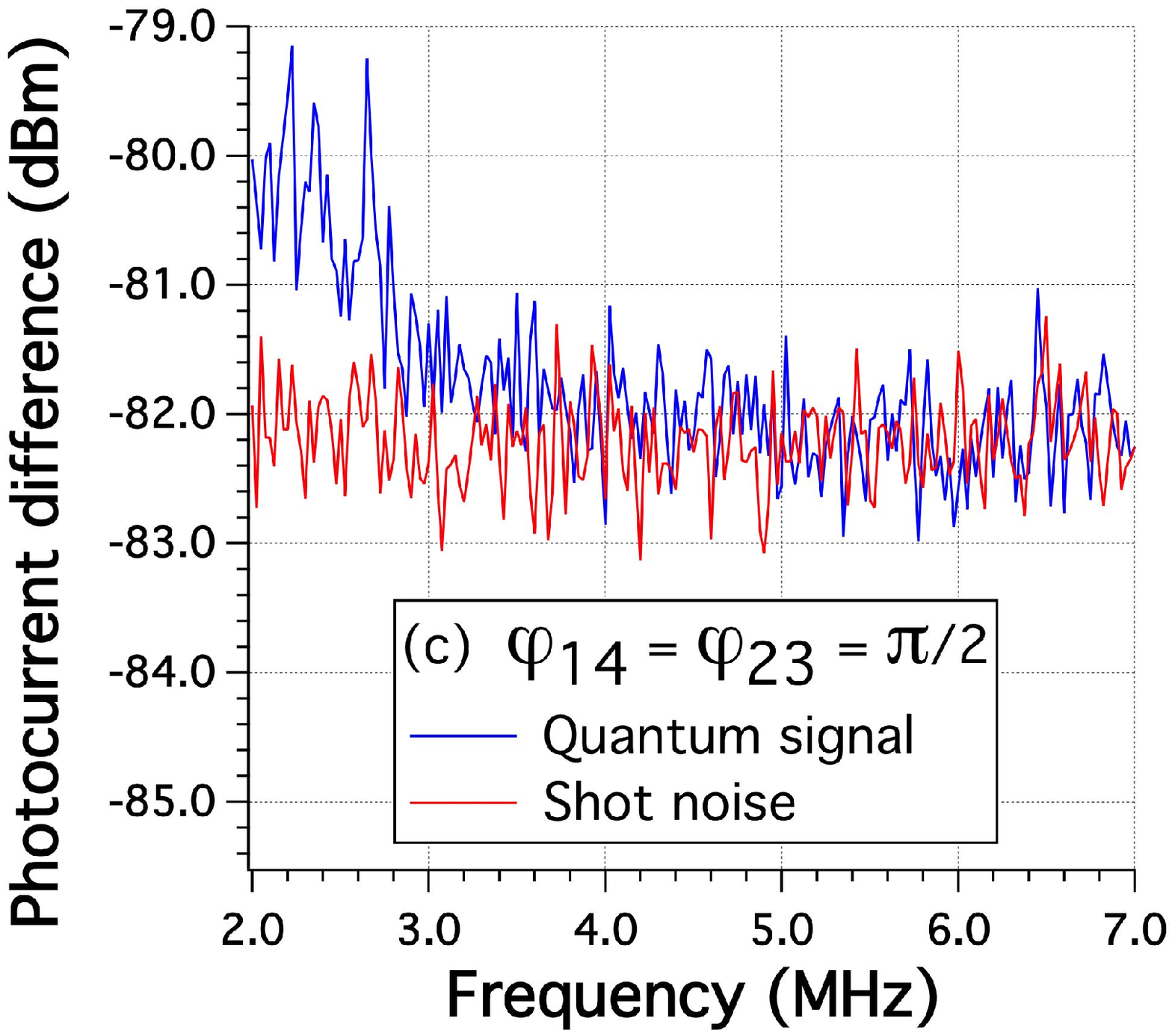}
}\parbox{0.5\columnwidth}{
\includegraphics[width=0.5\columnwidth]{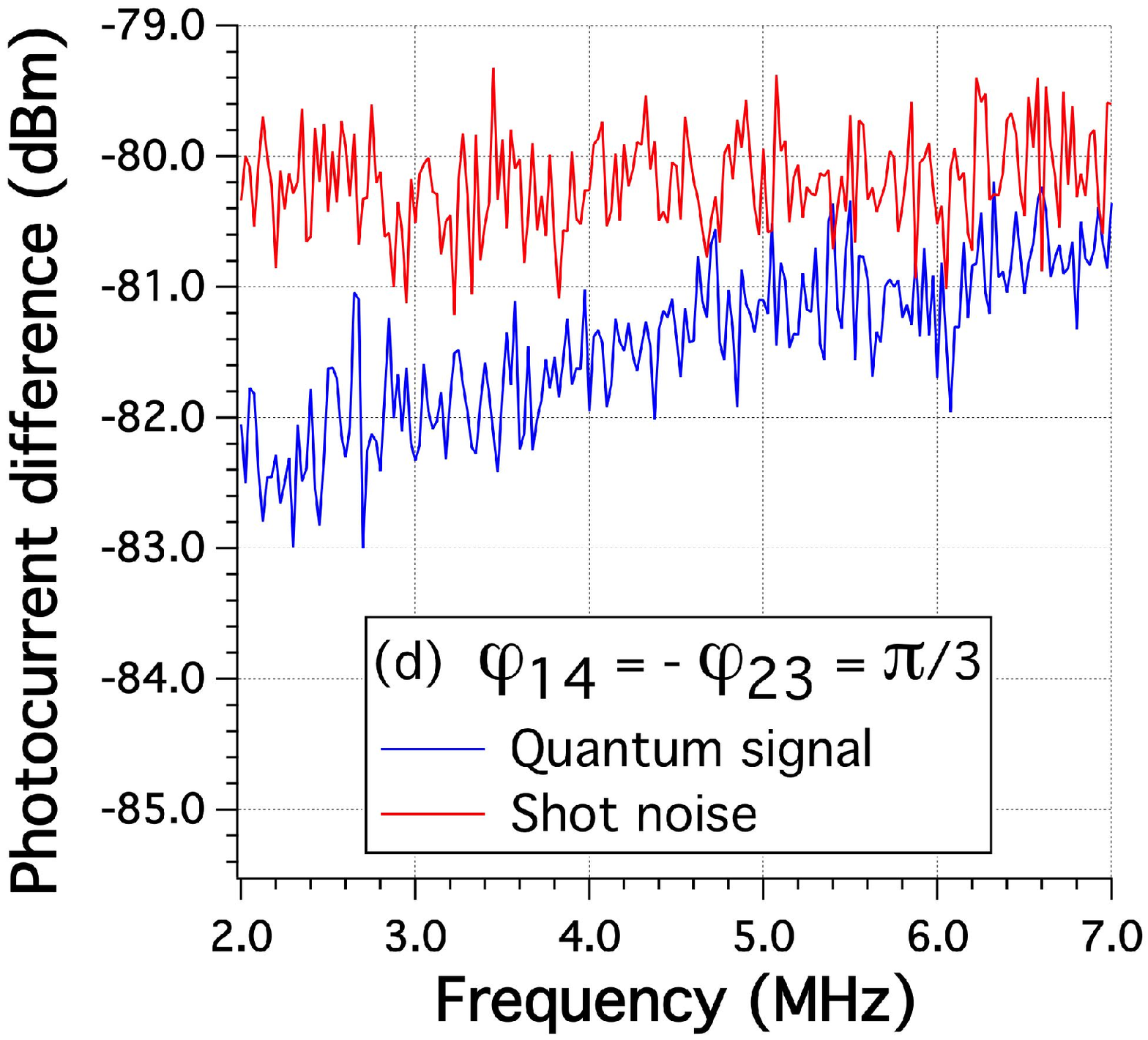}
}
\caption{Entanglement protection by injection of an independent EPR state into the beamsplitters. The output noise comprises: (a), amplitude-difference squeezing only, (b), amplitude-difference as well as phase-sum squeezing, (c), amplitude-difference squeezing and phase-difference {\em antisqueezing}, (d) hybrid squeezing case.  (Resolution bandwidth: 30 kHz. Video bandwidth: 10 kHz. 100 averages.)}
\label{fig:spec}
\end{center}
\end{figure}

\Fig{spec} illustrates the dramatic recovery of EPR correlations by measuring but a single output of each beamsplitter. 
Depending on the relative phase of the input fields at each beamsplitter, we obtained entanglement protection [\fig{spec}(a,b)] or failed to do so [\fig{spec}(c)], in complete agreement with \eq{dN}. Note that \fig{spec}(b) represents an indirect measurement of the EPR paradox as squeezing recovery is impossible to obtain in this case without initial simultaneous phase-sum and amplitude-difference squeezing. 

The lesser squeezing level in Fig. \ref{fig:spec}(b) can be explained by slight discrepancies in balancing out classical noise, and by the decrease of phase-sum squeezing above threshold in a doubly resonant OPO \cite{Reid1988} (although the OPO was never more than 4\% above threshold in all cases, which limited the degradation of phase-sum squeezing). Finally, the pump-induced variation of EPR amplitudes $\beta_{1,3}$ yielded no observable effect on decoherence protection, as predicted by the theory.

Note that no optical path was locked in the experiment, which prevented us from averaging the signal over long times. The values of phases $\varphi_{14, 23}$ were unambiguously determined, however, by use of the classical interference signals (\eq{class1}, \eq{class2}) which were monitored on low-frequency photodetection signals, the high-frequency spectrum being dedicated to quantum noise measurements.

\begin{figure}[hbt]
\begin{center}
\includegraphics[width=\columnwidth]{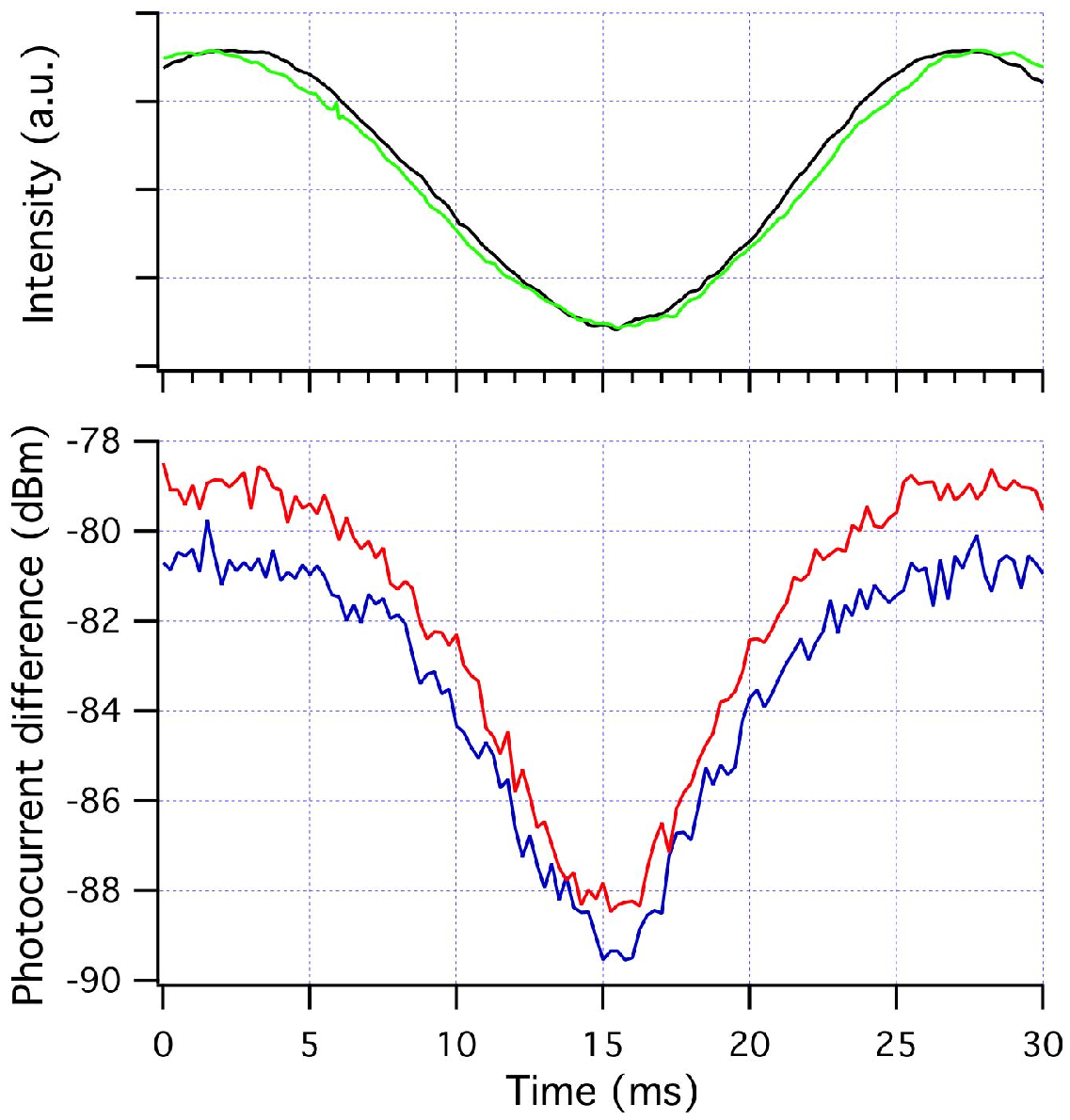}
\caption{Green, classical interference of $b_1$ and $b_4$ ($\cos\varphi_{14}$). Black, classical interference of $b_2$ and $b_3$ ($\cos\varphi_{23}$).  Fringe contrast for equal amplitudes was 95 \%. A slight chirp of the two PZT phase scans is visible but did not lead to large enough phase discrepancies over the scanning range (sensitivity to such discrepancies will increase with squeezing) and both phases can be considered equal and opposite for the whole scan. Blue, quantum signal $\av{[\delta(N_1\! -\! N_2)]^2}$. Red, shot noise. All traces were recorded synchronously. (Resolution bandwidth: 30 kHz. Video bandwidth: 1 kHz. 10 averages.)}
\label{fig:scan}
\end{center}
\end{figure}

Finally, \fig{spec}(d) confirms the prediction of \eq{final} for an arbitrary phase $\varphi=\pi/3$ that does {not} correspond to any EPR correlation in \eq{dN}. We then proceeded to test more thoroughly this phase independence of the squeezing by scanning $\varphi$. To do this, we performed zero-span spectral analysis at a frequency of 3 MHz and triggered all data acquisition to the synchronous scans of $\varphi_1$ and $\varphi_3$, which were tuned  as identically as possible so as to vary $\varphi_{\underline14}$ and $\varphi_{2 \underline3}$ in exact opposite directions. The results are displayed in Fig.\ref{fig:scan}.
They clearly show a preserved quantum noise reduction for all values of the relative phases satisfying $\varphi_{14}(t) = - \varphi_{23}(t)$.

In conclusion, we showed that the quantum amplitude correlations of entangled physical systems coupled to reservoirs are preserved if the reservoirs are themselves entangled. Note that the two EPR states are perfectly independent of each other, and the effect is also independent of their average respective amplitudes ($\beta_{1,3}$). The total squeezing is simply the average of the noise reductions in each EPR pair. Even though the initial EPR correlations were not fully measured (the amplitude-difference squeezing was observed but not the phase-sum), the theory shows that the observed restoration of amplitude-difference squeezing for certain phases is a purely nonlocal effect, impossible to obtain with locally squeezed inputs. The observation of this effect is therefore a confirmation of the emission of EPR correlated fields by the OPO. This system may also give rise to applications: Filip proposed its use to alleviating random couplings between few modes in optical fibers \cite{Filip2005}. It can also be viewed as a ``quantum tap'' for entanglement, since, by symmetry, an EPR state must be present at both outputs \cite{Olivares2009}, in a generalization of the single-mode case \cite{Bruckmeier1997}. In that case, the beamsplitters would be actual two-port-in, two-port-out devices rather than loss channels with one irretrievable output, and the setup could be used as a novel way to monitor and correct the effects of decoherence on an EPR state, with the help of another, stable EPR reference. Even though this protocol is EPR-state specific, this doesn't limit its applicability to protecting an EPR channel, say in a teleportation experiment \cite{Furusawa1998}.

We thank Stefano Olivares, Matteo Paris, and Gerd Leuchs for discussions. This work was supported by the U.S. National Science Foundation grants No.\ PHY-0960047 and No.\ PHY-0855632. 

\newpage



\end{document}